\documentclass[
superscriptaddress,
 amsmath,amssymb,
 aps,
reprint
]{revtex4-2}

\usepackage{color}
\usepackage{graphicx}
\usepackage{dcolumn}
\usepackage{bm}
\usepackage{epstopdf}
\usepackage{gensymb}

\usepackage{amsmath} 
\usepackage{booktabs}

\usepackage{mathptmx}

\usepackage{float}

\begin{document}

\title{Tunable polar distortions and magnetism in Gd$_x$La$_{1-x}$PtSb epitaxial films}
 
\author{Dongxue Du}
\affiliation{Materials Science and Engineering, University of Wisconsin-Madison, Madison, WI 53706}
\author{Chenyu Zhang}
\affiliation{Materials Science and Engineering, University of Wisconsin-Madison, Madison, WI 53706}
\author{Jingrui Wei}
\affiliation{Materials Science and Engineering, University of Wisconsin-Madison, Madison, WI 53706}

\author{Yujia Teng}
\affiliation{Department of Physics and Astronomy, Rutgers University}
\author{Konrad T. Genser}
\affiliation{Department of Physics and Astronomy, Rutgers University}

\author{Paul M. Voyles}
\affiliation{Materials Science and Engineering, University of Wisconsin-Madison, Madison, WI 53706}
\author{Karin M. Rabe}
\affiliation{Department of Physics and Astronomy, Rutgers University}
\author{Jason K. Kawasaki}
\affiliation{Materials Science and Engineering, University of Wisconsin-Madison, Madison, WI 53706}
\email{jkawasaki@wisc.edu}

\date{\today}
\begin{abstract}

Hexagonal $ABC$ intermetallics are predicted to have tunable ferroelectric, topological, and magnetic properties as a function of the polar buckling of $BC$ atomic planes. We report the impact of isovalent lanthanide substitution on the buckling, structural phase transitions, and electronic and magnetic properties of Gd$_x$La$_{1-x}$PtSb films grown by molecular beam epitaxy (MBE) on c-plane sapphire substrates. The Gd$_x$La$_{1-x}$PtSb films form a solid solution from x = 0 to 1 and retain the polar hexagonal structure ($P6_3 mc$) out to $x \leq 0.95$. With increasing $x$, the PtSb buckling increases and the out of plane lattice constant $c$ decreases due to the lanthanide contraction. While hexagonal LaPtSb is a highly conductive polar metal, the carrier density decreases with $x$ until an abrupt phase transition to a zero band overlap semimetal is found for cubic GdPtSb at $x=1$. The magnetic susceptibility peaks at small but finite $x$, which we attribute to Ruderman–Kittel–Kasuya–Yosida (RKKY) coupling between localized $4f$ moments, whose concentration increases with $x$, and free carriers that decrease with $x$. Samples with $x\geq 0.3$ show antiferromagnetic Curie-Weiss behavior and a Neel temperature that increases with $x$. The Gd$_x$La$_{1-x}$PtSb system provides opportunities to dramatically alter the polar buckling and concentration of local $4f$ moments, for tuning chiral spin textures and topological phases.

\end{abstract}

\maketitle

\section{Introduction}

Polar and inversion-symmetry breaking distortions are important for tuning properties of magnetic, ferroelectric, and topological materials. In particular, the polar hexagonal $ABC$ intermetallics with LiGaGe-type structure are predicted to have highly tunable properties as a function of the polar buckling in the $BC$ atomic planes. These properties include ferroelectricity \cite{bennett2012hexagonal}, hyperferroelectricity with persistent polarization down to the monolayer limit \cite{garrity2014hyperferroelectrics}, metal-insulator transitions \cite{genser2024first}, tunable Weyl and Dirac states \cite{gao2018dirac}, and Rashba states \cite{di2016intertwined}. 

In practice it remains challenging to tune the polar buckling. One set of approaches is to use the substrate, via epitaxial strain or substrate templating. With epitaxial strain, substrate-induced changes in lattice parameter can tune polar distortions. Examples include strain induced ferroelectricity in SrTiO$_3$ \cite{haeni2004room, schlom2007strain} and multiferroicity in EuTiO$_3$ \cite{lee2010strong, fennie2006magnetic}. For substrate templating, intra-unit cell distortions of the substrate directly template the desired distortion in the film. An example is tuning octahedral tilts in transition metal oxide films by selecting a substrate with the desired tilt pattern \cite{rondinelli2012control, yang2013untilting}. However, in these approaches the values of distortion are limited by the availability of substrates with the desired distortion pattern and lattice parameter. 

Free-standing membranes are an attractive alternative due to the ability to continuously apply homogeneous strains and strain gradients. Bending of membranes has been theoretically proposed as a route for switching polar metals \cite{zabalo2021switching} and for tuning chiral spin textures via the Dzyalshinskii-Moriya Interaction \cite{ga2022dzyaloshinskii}. Recent experiments on single crystalline membranes of GdAuGe \cite{laduca2024cold} and GdPtSb \cite{du2021epitaxy} demonstrate bending-induced changes in magnetic ordering, which may arise from changes in the AuGe or PtSb layer buckling. However, membrane control is still in its infancy and it remains challenging to control small radius of curvature bends repeatably and over large sample areas \cite{schmidt2001thin, kang2021pseudo, du2023strain}.

Here we use chemical pressure to continuously tune the polar buckling of $ABC$ intermetallic films grown by molecular beam epitaxy (MBE). Starting from LaPtSb, which is a high conductivity polar metal \cite{du2019high}, we show that substitution of Gd for La reduces the interlayer spacing and increases the PtSb layer buckling by a factor of 2, due to the smaller atomic radius of Gd. The buckling decreases the free carrier density. The magnetic susceptibility peaks for small finite $x$ and the apparent Neel temperature increases with $x$. We interpret the magnetic trends with $x$ in terms of RKKY coupling between local $4f$ moments, whose concentration increases with $x$, and the free carrier concentration that decreases with $x$.

\begin{figure*}[ht]
  \centering 
    \includegraphics[width=0.85\textwidth]{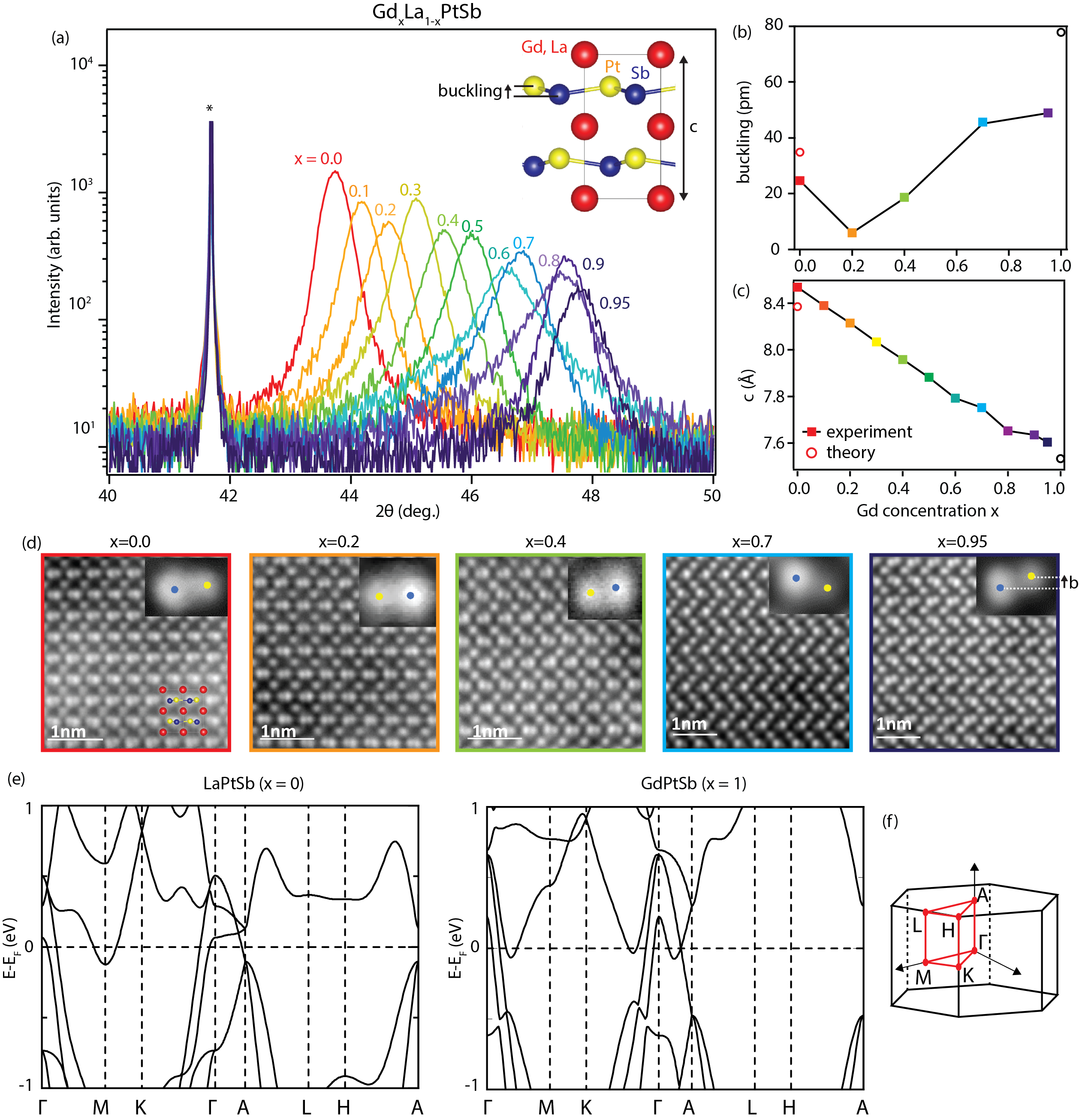}
    \caption{(a) Symmetric $2\theta-\omega$ x-ray diffraction scans of the $0004$ reflection. The asterisk marks an Al$_2$O$_3$ substrate reflection. Inset shows the crystal structure of $Ln$PtSb in the $P6_3 mc$ structure with polar buckling of the PtSb planes. (b) PtSb layer buckling determined from Gaussian intensity fitting of the STEM intensity (closed squares), compared to DFT calculated buckling in the $P6_3 mc$ structure (open circles). (c) Out of plane lattice parameter $c$ determined from x-ray diffraction (closed squares), compared to DFT calculated lattice parameter in the $P6_3 mc$ structure. (d) STEM images of samples with varying Gd composition $x$. (e) DFT-GGA bandstructures of relaxed $P6_3mc$ LaPtSb and GdPtSb without SOC. (f)  $P6_3mc$ Brillouin zone.}
    \label{structure}
\end{figure*}

\section{Results}

We first use density functional theory (DFT) calculations to predict the structural distortions with Gd substitution and the corresponding changes to electronic structure.  We focus on the end members LaPtSb and GdPtSb. Details of the computational methods are found in the Supplemental Materials.

\begin{figure*}[ht]
    \centering
    \includegraphics[width=0.8\textwidth]{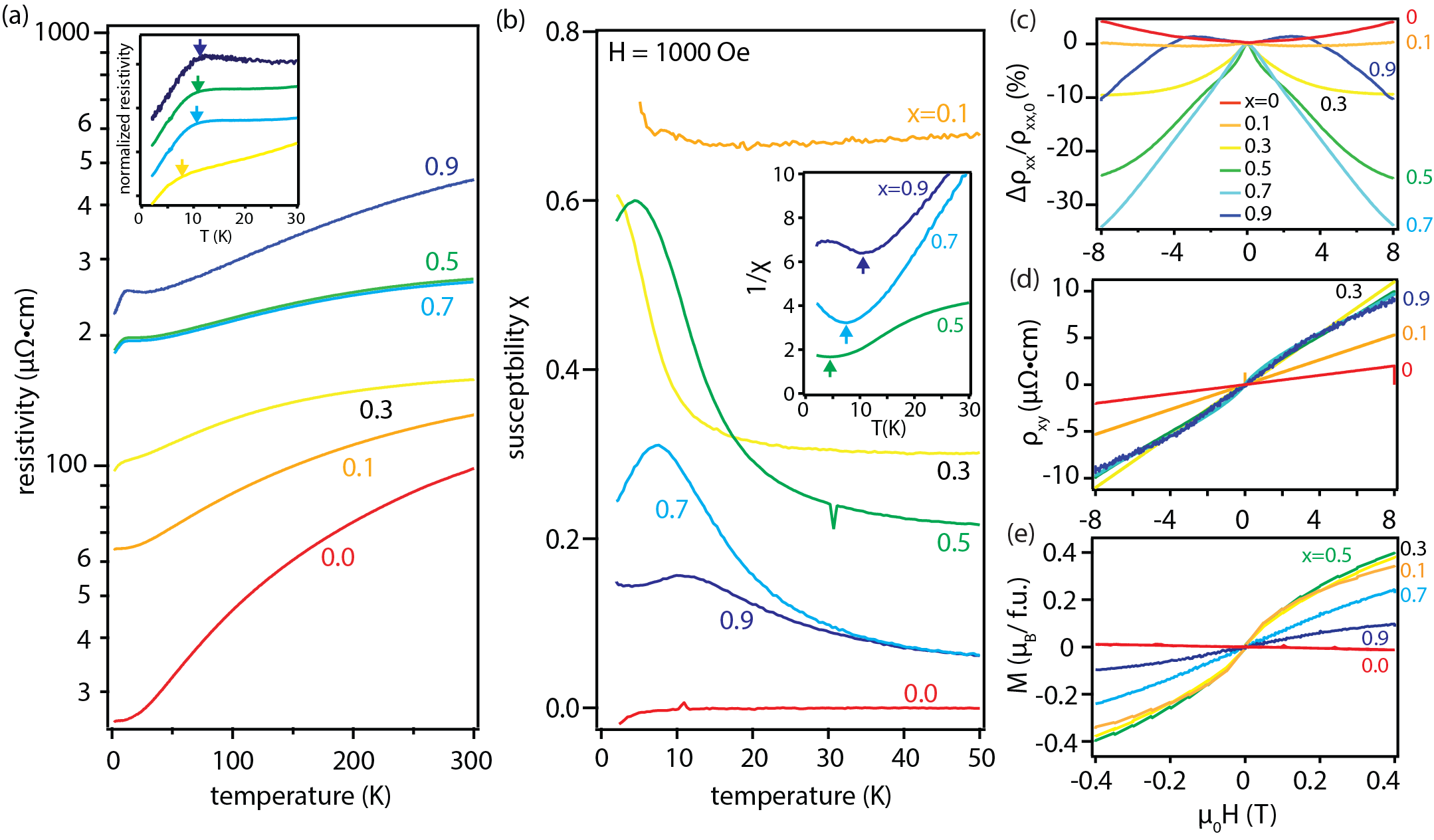}
    \caption{(a) Longitudinal resistivity versus temperature for films with varying Gd composition. Insert shows identification of $T^*$ via a tangent construction. (b) Magnetic susceptibility $\chi \approx M/H$ measured at 1000 Oe. (c) Longitudinal magnetoresistance versus applied field at 2 K. (d) Transverse (Hall) resistivity versus applied field at 2 K. (e) Magnetization versus applied field at 2 K.}
    \label{magnet_transport}
\end{figure*}

Fig. \ref{structure}(a) insert shows the crystal structure of LaPtSb, which experimentally crystallizes in the polar hexagonal LiGaGe-type structure (space group $P6_3 mc$) \cite{hoffmann2001alb2}. Here the Pt and Sb form a distorted wurtzite-like sublattice. The buckling of the Pt-Sb planes, which we define as the displacement of Pt and Sb atoms along the $c$ axis, breaks inversion symmetry and creates a unique polar axis along $c$. Our relaxed DFT calculations for LaPtSb predict a buckling of 35.1 pm and a $c$ lattice parameter of 8.37 \AA\ (Table 1). The corresponding electronic structure (without spin orbit coupling) is semimetallic with a band overlap of $\sim 1$ eV and primarily hole carriers (Fig. \ref{structure}(e)). 

GdPtSb experimentally crystallizes in the cubic half Heusler (space group F$\bar{4}$3m) with $abc-abc$ stacking \cite{du2023effect}, compared to the $ab-ab$ stacking of the hexagonal $P6_3 mc$ structure (Supplemental Fig. 1(c,d)). To assess the maximum buckling anticipated for Gd-substituted LaPtSb, we compute the relaxed structure for GdPtSb in a hypothetical $P6_3 mc$ structure. Our relaxed DFT calculations predict an enhanced buckling of 77.9 pm and a decreased $c$ of 7.533 \AA\ for GdPtSb (Supplemental Table 1). The corresponding electronic structure shows a smaller band overlaps than LaPtSb, including an increased band gap at $A$ and an increase in the conduction band minima at $M$ and $H$ (Fig. \ref{structure}(e)). 

Wide angle $2\theta-\omega$ XRD scans reveal that our MBE-grown Gd$_x$La$_{1-x}$PtSb films form a continuous solid solution from x = 0 to 1 with no secondary phases and retain the hexagonal structure out to $x \leq 0.95$ (Fig. \ref{structure}(a) and Supplemental Fig. 1(a)). With increasing Gd composition $x$ we observe an increase in the Bragg angle of the hexagonal $0004$ reflection, corresponding to a decreasing out of plane lattice constant $c$ (Fig. \ref{structure}(c), filled squares). This experimental trend is in good agreement with our DFT calculations for LaPtSb and GdPtSb (Fig. \ref{structure}(c) open circles and Table 1). We attribute the decreased $c$ to the smaller ionic radius of Gd compared to La, due to the lanthanide contraction. The in-plane lattice parameters, extracted from off axis scans of the $10\bar{1}2$ reflections, show a much weaker dependence on $x$ as they vary from 4.50 to 4.58 \AA\ for $x\leq 0.95$. Details of MBE synthesis and XRD characterization appear in Supplemental Materials.

We use off axis azimuthal $\phi$ scans to distinguish cubic from hexagonal polymorphs. We find that the hexagonal structure is retained out to $x \leq 0.95$. These samples display a six-fold rotation of the $10\bar{1}2$ reflection corresponding to the hexagonal structure, whereas the $x=1$ end member GdPtSb displays a three-fold rotation corresponding to cubic structure (Supplemental Fig. 1(b)). The hexagonal $10\bar{1}2$ film reflections for $x \leq 0.95$ align in $\phi$ with the $10\bar{1}4$ reflections of the sapphire substrates, corresponding to an epitaxial relationship of Gd$_x$La$_{1-x}$PtSb $[1 0 \bar{1} 0] (0001) \parallel$ Al$_2$O$_3$ $[1 0 \bar{1} 0] (0001)$. For the $x=1$ cubic sample, in addition to the three-fold $220$ reflections that align with the sapphire $10\bar{1}4$, we observe a second set of weaker reflection that are rotated in $\phi$ by 60 degrees. We attribute this second set to an antiphase domain.

We quantify the buckling using high angle annular diffraction (HAADF) scanning transmission microscopy (STEM imaging, see Supplemental Materials). In Fig. \ref{structure}(d) the brightest atomic columns correspond to Pt, next brightest to Gd/La, and lowest intensity to Sb. We observe hexagonal $ab-ab$ layer stacking for $x\leq 0.95$ (Fig. \ref{structure}(d)) and cubic $abc-abc$ stacking for $x=1$  (Supplemental Fig. 1(d)), consistent with the XRD $\phi$ scans. From 2D Gaussian fitting of the intensity we extract the polar buckling. We find an overall increase in the buckling with Gd composition $x$, consist with our DFT calculations, which we attribute to stronger interlayer bonding with decreased $c$. We note, however, that measured bucklings are $\sim 40\%$ smaller than the DFT predicted bucklings.

Temperature dependent resistivity $\rho$ reveals a metal-like temperature dependence $d\rho/dT>0$ for all samples and an increase in $\rho$ with Gd composition (Fig. \ref{magnet_transport}(a)). This is consistent with our DFT calculations that predict decreased band overlaps for GdPtSb compared with LaPtSb (Fig. \ref{structure}(e)). Samples with intermediate Gd composition $0.3 \leq x \leq 0.9$ display kinks in the resistivity versus temperature at a temperature $T^*$, which we define by a local tangent construction (Supplemental Fig. 2). We attribute these kinks to magnetic phase transitions. The kinks correspond approximately, but not exactly, to kinks that we observe in the inverse of the magnetic susceptibility versus temperature (Fig. \ref{magnet_transport}(b) insert). 

The 2 K magnetoresistance is shown in Fig. \ref{magnet_transport}(c). Samples with $x\leq 0.1$ show a weak positive quadratic dependence on applied field. With increasing $x$ we observe a more negative magnetoresistance, which we attribute to field alignment of $4f$ moments that suppresses free carrier scattering. Alternatively, a large negative magnetoresistance can arise from the chiral anomaly in Weyl semimetals \cite{hirschberger2016chiral}, and many materials within the $ABC$ hexagonal family are predicted to have Weyl points that are tunable with layer buckling \cite{di2016intertwined, gao2018dirac}. More systematic angular dependent magnetoresistance measurements and test against trivial forms of scattering are required to identify whether the negative magnetoresistance arises from chiral anomaly \cite{du2023effect, hirschberger2016chiral}.

Hall effect measurements (Fig. \ref{magnet_transport}(d)) show a positive sign indicating hole dominated transport. Samples with $x\leq 0.1$ show a linear $\rho_{xy}$ vs applied field and are well fit to model with a single hole band. Samples with $x>0.1$ show nonlinear behavior at low fields $\mu_0 H < 4$ T, which may arise from an anomalous Hall effect consistent with the nonlinear $M(H)$ measured by SQUID (Fig. \ref{magnet_transport}(e)). For these samples we extract a hole density from a high field fit to the linear component of $\rho_{xy}(B)$ (Supplemental Fig. 3). The results of the Hall modelling are shown in Fig. \ref{summary}(c), where we find a decrease in the hole density with increasing Gd composition. We attribute this decreasing carrier density to the increased Pt-Sb layer buckling, which decreases the bandwidths and band overlaps at the Fermi energy. The Gd $4f$ bands, which are located $\sim 8$ eV below the Fermi energy by DFT calculations, are not anticipated to contribute free carriers.

SQUID magnetometry measurements at 2 K show that LaPtSb is diamagnetic with a negative and linear $M$ vs applied field $H$ (Fig. \ref{magnet_transport}(e)). For $0.1 \leq x \leq 0.5$, we observe a positive nonlinear dependence of $M \propto \tanh(H)$ with no clear hysteresis, suggestive of superparagmagnetism from dilute Gd $4f$ moments or clusters of $4f$ moments. For $x\geq 0.7$ we observe a linear positive $M(H)$. Fig. \ref{magnet_transport}(b) shows the temperature dependent magnetic susceptibility, which we define as $M/H$ at low field H = 1000 Oe. For samples with $x \geq 0.5$ we observe a negative Curie-Weiss temperature (Fig. \ref{magnet_transport}(b) insert), consistent with antiferromagnetic transitions. The apparent Neel transition temperature versus Gd composition is shown in Fig. \ref{summary}(a). Whereas GdPtSb is anticipated to be a G-type antiferromagnet \cite{du2021epitaxy} consistent with the ordering of GdPtBi \cite{suzuki2016large}, the spin structure of Gd$_x$La$_{1-x}$PtSb for $x<1$ is not known and requires further study.

\section{Discussion}

\begin{figure}[h!]
    \centering
    \includegraphics[width=0.3\textwidth]{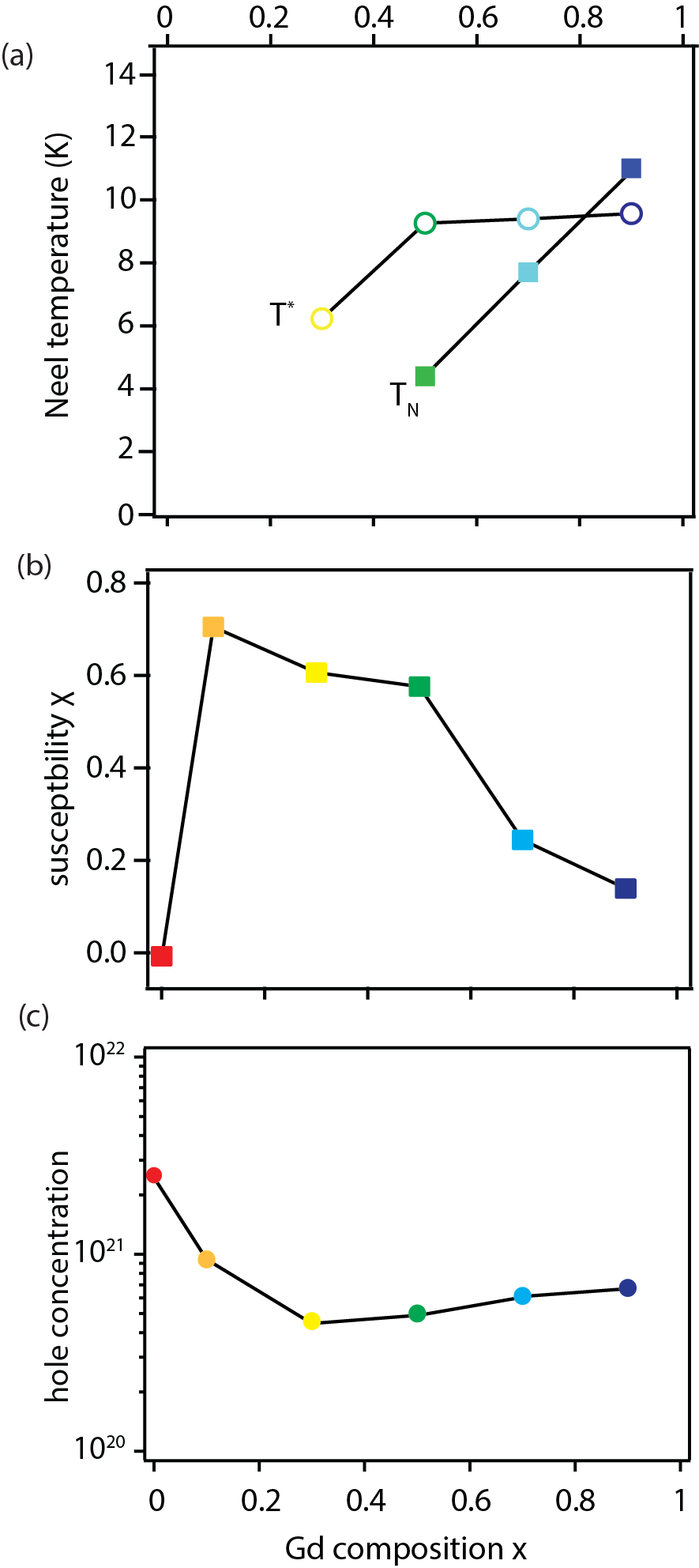}
    \caption{(a) Neel temperature $T_N$ extracted from kink in $1/\chi$ versus $T$ (filled squares) and $T^*$ extracted from the knink in $\rho(T)$ (open circles). (b) Magnetic susceptibility at 2 K, defined as $M/H$ at $ H = 1000$ Oe. (c) Carrier concentration extracted from Hall effect using a single band model.}
    \label{summary}
\end{figure}

We summarize the key trends of magnetism and transport with Gd composition $x$ in Fig. \ref{summary}. Fig. \ref{summary}(a) plots the characteristic temperatures: $T^*$ from kink in $\rho(T)$ and the Neel temperature $T_N$ from the kink in the inverse of magnetic susceptibility versus temperature. Both characteristic temperatures increase monotonically with Gd composition, which we attribute to the increasing concentration of Gd $4f$ moments. We find that $T_N$ and $T^*$ are close, but do not exactly coincide. The difference in $T^*$ and $T_N$ may be due to an intermediate magnetic state, since other lanthanide-containing $ABC$ compounds display multiple transitions due to magnetic frustration \cite{ram2023multiple}. 

We caution, however, that instead of kinks in $\rho(T)$, magnetic transitions observed by transport measurements are more formally related to anomalies in the derivative $d\rho/dT$ \cite{fisher1968resistive}. In particular, in the limit of small Fermi wavenumber $k_F$ the magnetic transition temperature corresponds to the peak in $d\rho/dT$ \cite{fisher1968resistive}. However, this simple relationship rapidly breaks down for larger $k_F$. Our DFT calculations suggest that hexagonal Gd$_x$La$_{1-x}$PtSb has large $k_F$ (Fig. \ref{structure}(e)), consistent with the large carrier density that we observe by Hall effect (Fig. \ref{magnet_transport}(d) and Fig. \ref{summary}). The resulting experimental $d\rho/dT$ curves display broad minima in which it is difficult to define a clear transition temperature (Supplemental Fig. 2). Therefor in our data we find that the tangent construction of $T^*$ provides a more clearly defined transition.


The hole concentration extracted from Hall effect at 2 K decreases with Gd composition (Fig. \ref{summary}c). We attribute this trend to the decrease in PtSb layer buckling with $x$, which decreases the intralayer hoppings and decreases the band overlap at the Fermi energy.

The magnetic susceptibility $\chi$ dependence on $x$ is more complex: starting from a small negative value for diamagnetic LaPtSb at $x=0$, the susceptibility rapidly peaks for $x=0.1$ and then gradually decreases (Fig. \ref{summary}b). We interpret the non monotonic trend in $\chi$ to arise from the Ruderman–Kittel–Kasuya–Yosida (RKKY) interaction \cite{ruderman1954indirect, kasuya1956theory} in which local Gd $4f$ are coupled via the Fermi sea. With increasing $x$ the concentration of Gd $4f$ moments increases; however, the concentration of hole carrier decreases due to the layer buckling. This causes the peak susceptibility to occur at an intermediate Gd concentration. The peak in $\chi(x)$ corresponds to the samples that show superparamagnetic $M(H)$ at low Gd concentration $(x \leq 0.5)$. For larger values of Gd concentration $x> 0.5$, as the Gd moments are closer together, the RKKY coupling is antiferromagnetic resulting in a linear $M(H)$ and a decrease in susceptibility.

In conclusion, we demonstrated epitaxial synthesis of Gd$_x$La$_{1-x}$PtSb films on Al$_2$O$_3$ (0001) with highly tunable polar buckling that is 2 times larger than the buckling for LaPtSb. The hole concentration decreases with Gd-induced layer buckling. Gd substitution also contributes local $4f$ moments, which order antiferromagnetically for high concentrations $x$. We explain the magnetic susceptibility in terms of a competition between Pt-Sb buckling that decreases the hole concentration, versus an increase in the concentration of Gd $4f$ moments in an RKKY picture. The lanthanide substituted $ABC$ intermetallics offer a continuously tunable platform for exploring the effects of polar distortions on magnetism in a $4f$ system.

\section{Acknowledgment}

Transport and magnetometry by JKK and DD were supported by the Air Force Office of Scientific Research (FA9550-21-0127) and by the Army Research Office (W911NF-17-1-0254). Epitaxial synthesis by DD and JKK was supported by the U.S. Department of Energy (DE-SC0023958). Data analysis and writing of the manuscript by all authors were supported in part by the National Science Foundation via the Wisconsin Materials Research Science and Engineering Center (MRSEC, DMR-2309000). DFT calculations by KMR, KG, and YT were supported by ONR grant N00014-21-1-2107. STEM measurements by CZ, JW, and PMV were supported by the NSF through the University of Wisconsin Materials Research Science and Engineering Center (DMR-2309000). We gratefully acknowledge use of facilities and instrumentation supported by NSF through the University of Wisconsin Materials Research Science and Engineering Center (DMR-2309000).

\bibliographystyle{apsrev}
\bibliography{ref}

\end{document}